\documentclass[%
 aip,cha,twocolumn
 sd,%
 reprint,%
]{revtex4-1}

\usepackage{graphicx}% Include figure files
\usepackage{dcolumn}% Align table columns on decimal point
\usepackage{bm}% bold math
\usepackage{amsfonts}
\usepackage{eurosym}
\usepackage{amssymb}
\usepackage{amsmath}
\usepackage[dvips]{epsfig}
\usepackage{color}
\usepackage{epstopdf}
\usepackage{ntheorem}
\usepackage{subfig}

\theoremseparator{.}
\newtheorem{ntheorem}{Theorem}

\begin{document}

%\preprint{AIP/123-QED}

\title{Stable dipole solitons and soliton complexes in the nonlinear
Schr\"{o}dinger equation with periodically modulated nonlinearity}

\author{M. E. Lebedev}
\author{G. L. Alfimov}
\affiliation{National Research University of Electronic Technology MIET, Zelenograd, Moscow 124498, Russia}
\email{gloriouslair@gmail.com, galfimov@yahoo.com.}

\author{Boris A. Malomed}
\affiliation{Department of Physical Electronics, School of Electrical
Engineering, Faculty of Engineering, Tel Aviv University, Tel Aviv 69978,
Israel}
\email{malomed@post.tau.ac.il.}

\date{\today}% It is always \today, today,

\begin{abstract}
We develop a general classification of the infinite number of families of
solitons and soliton complexes in the one-dimensional
Gross-Pitaevskii/nonlinear Schr\"{o}dinger equation with a nonlinear lattice
\textit{pseudopotential}, i.e., periodically modulated coefficient in front
of the cubic term, which takes both positive and negative local values. This
model finds direct implementations in atomic Bose-Einstein condensates and
nonlinear optics. The most essential finding is the existence of two
branches of \textit{dipole solitons }(DSs), which feature an antisymmetric
shape, essentially squeezed into a single cell of the nonlinear lattice. 
This soliton species was not previously considered in nonlinear lattices.
We demonstrate that one branch of the DS family (namely, the one which
obeys the Vakhitov-Kolokolov criterion) is\emph{\ stable}, while unstable
DSs spontaneously transform into stable fundamental solitons (FSs). The
results are obtained in numerical and approximate analytical forms, the
latter based on the variational approximation. Some stable bound states of
FSs are found too.
\end{abstract}

\maketitle

\begin{quotation}
Periodic (alias lattice) potentials is a well-known ingredient of
diverse physical settings represented by the nonlinear 
Schr\"{o}dinger/Gross-Pitaevskii equations. The lattice potentials help to create
self-trapped modes (solitons) which do not exist otherwise, or stabilize
those solitons which are definitely unstable in free space. In particular,
the lattice potentials generate the bandgap spectrum in the linearized
version of the equation, and adding local cubic nonlinearity gives rise to a
great variety of \textit{gap solitons} and their bound complexes residing in
the spectral gaps. On the other hand, an essential extension of the
concept of lattice potentials is the introduction of nonlinear \textit{%
pseudopotentials}, which are induced by spatially periodic modulation of the
coefficient in front of the cubic term. While single-peak fundamental  
solitons (FSs) in nonlinear potentials were studied in detail, more sophisticated ones,
such as narrow antisymmetric \textit{dipole solitons} (DSs), which essentially  
reside in a single cell of the nonlinear lattice, were not previously 
considered in this setting. Their shape is similar to that of the so-called 
\textit{subfundamental} species of gap solitons in linear lattices,
which have a small stability region. In this work, we first develop a general 
classification of a potentially infinite number of different types of soliton 
complexes supported by the nonlinear lattice. For physical applications, the most 
significant finding is the existence of two branches of the DS family, one of which is 
entirely \emph{stable}. Its stability is readily predicted by the celebrated
Vakhitov-Kolokolov criterion, while the shape of the branch is qualitatively
correctly predicted in an analytical form by means of the variational approximation. In
addition to that, it is found that some bound states of FSs are stable too,
although a majority of such complexes are unstable.
\end{quotation}

%\documentclass[aip,cha,reprint,twocolumn]{revtex4-1}

%\begin{document}

\maketitle

\section{Introduction}

\label{Intro}It is well known that the variety of bright solitons, supported
by the balance between the self-focusing nonlinearity and diffraction (in
optics) or kinetic energy (in matter waves), can be greatly expanded if a
spatially periodic (alias lattice) potential is introduced, in the form of
photonic lattices acting on optical waves \cite{phot-latt}, or optical
lattices acting on matter waves in atomic Bose-Einstein condensates (BECs)
\cite{opt-latt}. In particular, periodic potentials make it possible to
create \textit{gap solitons} in media with self-defocusing nonlinearity, due
to its interplay with the effective negative mass of collective excitations,
see original works \cite{Dark}-\cite{Oberthaler} and books \cite{JYang,DPeli}%
. In addition to the fundamental solitons, the analysis addressed patterns
such as nonlinear Bloch states \cite{Kiv2003,BlochW}, domain walls \cite%
{DomWalls05}, and gap waves, i.e., broad modes with sharp edges \cite%
{gapwave}.

The spectral bandgap structure induced by lattice potentials gives rise to
many families of gap solitons, classified by the number of a bandgap in
which they reside. Further, the oscillatory shape of fundamental gap
solitons opens the way to build various two- and multi-soliton bound states
through the effective interaction potential induced by their overlapping
tails. The variety of the gap-soliton families include both stable and
unstable solutions. A specific possibility, revealed in work \cite{Thaw} and
further analyzed in Refs. \cite{sub}-\cite{China}, is the existence of
\textit{subfundamental solitons} (SFSs) in the second finite bandgap (in Ref. 
\cite{China} SFSs were called ``second-family fundamental gap solitons"). They
feature a dipole (antisymmetric) shape, which is squeezed, essentially, into
a single cell of the lattice potential. The name ``subfundamental'' implies that the
soliton's norm (in terms of BEC; or the total power, in terms of optics)
is smaller than the norm of a stable fundamental soliton (FS) existing at
the same value of the chemical potential (or propagation constant, in the
optics model) in the second finite bandgap. SFSs have a small stability
region \cite{China}, while unstable ones spontaneously rearrange 
into stable FSs belonging to the first
finite bandgap. Partial stabilization of SFSs was also demonstrated in a model
which includes, in addition to the local nonlinearity, long-range
dipole-dipole interactions \cite{subDD}.

Apart from the linear spatially periodic potentials induced by lattice
structures, the formation of solitons may be facilitated by
nonlinear-lattice \textit{pseudopotentials} \cite{pseudo}, which are induced
by spatially periodic modulation of the coefficient in front of the cubic
term in the respective Gross-Pitaevskii/nonlinear Schr\"{o}dinger equation
(GPE/NLSE) \cite{RMP}. This structure can be created in BEC by means of the
Feshbach resonance controlled by magnetic or optical fields \cite{FR-Randy}-%
\cite{FR-Tom}. Experimentally, the possibility of the periodic modulation of
the nonlinearity on a submicron scale was demonstrated in Ref. \cite%
{experiment-inhom-Feshbach}. The spatial profile of the nonlinearity may
also be ``painted" by a fast moving laser beam \cite{painting}, or imposed
by an optical-flux lattice \cite{Cooper}. Another approach relies on the use
of a magnetic lattice, into which the atomic condensate is loaded \cite%
{magn-latt}, or of concentrators of the magnetic field \cite{concentrator}.
In optics, spatial modulation of the Kerr coefficient can be achieved by
means of an inhomogeneous density of resonant nonlinearity-enhancing dopants
implanted into the waveguide \cite{Kip}. Alternatively, a spatially periodic
distribution of resonance detuning can be superimposed on a uniform dopant
density. A review of results for solitons supported by nonlinear lattices
was given in Ref. \cite{RMP}.

In the one-dimensional setting, a generic form of the scaled GPE/NLSE for
the mean-field amplitude, $\Psi (x,t)$, including both a linear periodic
potential, $U(x)$, and a periodic pseudopotential induced by modulation
function $P(x)$, both with period $L$, is \cite{HS}
\begin{equation}
i\Psi _{t}+\Psi _{xx}-U(x)\Psi +P(x)|\Psi |^{2}\Psi =0.  \label{UP}
\end{equation}%
The prototypical examples of both periodic potentials are provided by
functions%
\begin{equation}
\left\{ U(x),P(x)\right\} =\left\{ A_{U},A_{P}\right\} +\left\{
B_{U},B_{P}\right\} \cos (2x),  \label{cos}
\end{equation}%
where the period is scaled to be $L=\pi $. Equation (\ref{UP}) is written in
terms of BEC; in optics, Eq. (\ref{UP}) models the light propagation in
planar waveguides, with transverse coordinate $x$, $t$ being replaced by the
propagation distance, $z$. In the former case, the model can be implemented
in a cigar-shaped BEC trap with the transverse confinement strength subject
to periodic modulation along the axial direction, $x$ \cite{De Nicola,Luca}.
Similarly, the optics realization is possible in the planar waveguides with
the thickness (in direction $y$) subject to the same modulation along $x$.
It is also relevant to mention that, while we here consider the simplest
cubic form of the local nonlinearity in Eq. (\ref{UP}), strong transverse
confinement applied to the BEC with a relatively high atomic density gives
rise to the one-dimensional equation with nonpolynomial nonlinearity \cite%
{Luca}, which may be a subject for a separate work. \textcolor{black}{It is important for the what follows that Eq.~(\ref{UP}) conserves the quantities $N$ and $E$,
\begin{eqnarray}
&&N=\int_{-\infty}^\infty |\Psi|^2~dx,\label{N_cons}\\[2mm]
&&E = \int \limits_{-\infty}^{+\infty}\left( | \Psi_x |^2 +U(x) | \Psi |^2 -\frac12 P(x) | \Psi |^4\right)~dx \label{E_cons}
\end{eqnarray}
having in BEC context the sence of the number of particles and the energy correspondingly.}

The objective of the present work is to generate new types of solitons in
the model based on Eq. (\ref{UP}), and identify \emph{stable solitons} among
them. To this end, we develop a procedure which makes it possible to predict
an infinite number of different families of stationary soliton solutions
(starting from the SF and DS families), by means of a \textit{coding
technique }\cite{AlfAvr}. Actual results are produced, with the help of
numerical calculations, for the model with the pseudopotential only \cite%
{Malomed}, i.e., Eq. (\ref{UP}) with $U=0$, where effects produced by the
periodic modulation of the nonlinearity are not obscured by the
linear-lattice potential. Keeping in mind the prototypical $\cos (2x)$
modulation function in Eq. (\ref{cos}), we assume that $P(x)$ in Eq. (\ref%
{UP}) is an even $\pi $-periodic function, which takes both positive and
negative local values. In particular, while FSs supported by nonlinear
lattices have been already studied in detail \cite{Malomed}, a possibility
of the existence and stability of the single-cell DSs in the same setting was not
considered previously. We demonstrate that this class of solitons is also
supported by the nonlinear lattice. It is composed of two branches, one of
which is \emph{stable}, on the contrary to the chiefly unstable SFS
family in the models with linear lattices. Another difference is that 
the single-cell DSs are not subfundamental, as their norm exceeds that 
of the SFs existing at the same value of the soliton frequency. We also show that, in addition to
the SFs and DSs, there exist a plethora of solitons in the model with
periodic pseudopotential. While most of them are unstable, we have found
some stable bound states of fundamental solitons.

The rest of the paper is structured as follows. Stationary soliton solutions
are produced in Section II. Results of the stability analysis are summarized
in Section III. Section IV is focused on the new class of the single-cell DSs,
including both numerical results and analytical approximations, based on the
variational approximation (VA) and Vakhitov-Kolokolov (VK) \cite{Vakh}
stability criterion. The paper is concluded by Section V.

\section{Stationary modes}

\label{StatMod}

Stationary solutions to Eq. (\ref{UP}) with chemical potential $\omega $ (in
the optics model, $-\omega $ is the propagation constant) are sought for in
the usual form, $\Psi (t,x)=u(x)\exp \left( -i\omega t\right) $, where $u(x)$
is determined by equation
\begin{equation}
u_{xx}+Q(x)u+P(x)u^{3}=0,\quad Q(x)=\omega -U(x).
\label{eq:stationary}
\end{equation}%
Solitons are selected by the localization
condition,
\begin{equation}
\lim\limits_{x\rightarrow \pm \infty }u(x)=0,  \label{Local}
\end{equation}%
which implies that the function $u(x)$ is real (see, e.g., Ref. \cite{AKS}).
Therefore, we focus our attention on real solutions to Eq. (\ref%
{eq:stationary}).

For the analysis of stationary modes we apply the approach developed
previously for the usual model, with the uniform nonlinearity and a linear
lattice potential, i.e., $P(x)=-1$ and $U(x)$ a bounded periodic function
\cite{AlfAvr}. This approach makes use of the fact that the ``most common"
solutions of equation
\begin{equation}
u_{xx}+Q(x)u-u^{3}=0  \label{eq:old}
\end{equation}%
are singular, i.e., they diverge at some finite value of $x = x_0$ ($\lim
\limits_{x \to x_0} u(x) = \infty$), as
\begin{equation}
u(x)\approx \pm \sqrt{2}\left( x-x_{0}\right) ^{-1}.  \label{diverging}
\end{equation}

Then, it was shown that, under certain conditions imposed on $Q(x)$,
nonsingular solutions can be described using methods of symbolic dynamics.
More precisely, under these conditions there exists one-to-one
correspondence between all solutions of Eq. (\ref{eq:old}) and bi-infinite
sequences of symbols of some finite alphabet, which are called \textit{codes}
of the solutions.

As shown below, this approach can be extended for Eq.~(\ref{eq:stationary}),
which combines the periodic lattice potential and periodic modulation of the
nonlinearity coefficient, that represents the nonlinear-lattice
pseudopotential.

\subsection{The coding procedure}

\label{Coding}

Assume that $Q(x)$ and $P(x)$ in Eq. (\ref{eq:stationary}) are even $\pi $%
-periodic functions. We call a solution $u(x)$ of Eq.(\ref{eq:stationary})
singular if $u(x)$ diverges at finite $x_{0}$ as per Eq. (\ref{diverging}).
In this case, one may also say that solution $u(x)$ \textit{collapses} at
point $x=x_{0}$.

Define \textit{Poincar\'{e} map} $T:\mathbb{R}^{2}\rightarrow \mathbb{R}^{2}$
associated with Eq.(\ref{eq:stationary}) as follows:
\begin{equation}
T%
\begin{pmatrix}
u_{0} \\
u_{0}^{\prime }%
\end{pmatrix}%
=%
\begin{pmatrix}
u(\pi ) \\
u_{x}(\pi )%
\end{pmatrix}%
\end{equation}%
where $u(x)$ is a solution of the Cauchy problem for Eq. (\ref{eq:stationary}%
) with initial conditions%
\begin{equation}
u(0)=u_{0},\quad u_x(0)=u_{0}^{\prime }.  \label{initial}
\end{equation}
We call an \textit{orbit} a sequence of points $\{p_{n}\}$, $p_{n}\in
\mathbb{R}^{2}$ (the sequence may be finite, infinite or bi-infinite), such
that $Tp_{n}=p_{n+1}$.

Define sets $\mathcal{U}_{L}^{+}\in \mathbb{R}^{2}$ and $\mathcal{U}%
_{L}^{-}\in \mathbb{R}^{2}$, $L>0$ as follows: $p=(u_{0},u_{0}^{\prime })\in
\mathcal{U}_{L}^{+}$ if and only if solutions of the Cauchy problem for Eq. (%
\ref{eq:stationary}) with initial conditions (\ref{initial}) does not
collapse on interval $[0,L]$. Similarly, we define $\mathcal{U}_{L}^{-}$ as
the set of initial conditions $u(0)=u_{0}$, $u_{x}(0)=u_{0}^{\prime }$ such
that the corresponding solution of the Cauchy problem for Eq.(\ref%
{eq:stationary}) does not collapse on interval $[-L,0]$. It is easy to show
that Poincar\'{e} map $T$ is defined only on set $\mathcal{U}_{\pi }^{+}$
and transforms it into $\mathcal{U}_{\pi }^{-}$. Accordingly, inverse map $%
T^{-1}$ is defined only on $\mathcal{U}_{\pi }^{-}$ and transforms this set
into {\color{black} $\mathcal{U}_{\pi }^{+}$}.

Next, consider the following sets:
\begin{eqnarray}
&&\Delta _{0}=\mathcal{U}_{\pi }^{+}\cap \mathcal{U}_{\pi }^{-}, \\
&&\Delta _{n+1}^{+}=T\Delta _{n}^{+}\cap \Delta _{0},\quad n=0,1,\ldots , \\
&&\Delta _{n+1}^{-}=T^{-1}\Delta _{n}^{-}\cap \Delta _{0},\quad n=0,1,\ldots
,
\end{eqnarray}%
Evidently, $\Delta _{0}$ consists of points that have $T$-image and $T$%
-pre-image. The following statements are valid:
\begin{eqnarray}
&&\{p\in \Delta _{n}^{+}\}\quad \Leftrightarrow \quad \{Tp,T^{2}p,\ldots
,T^{n}p\in \Delta _{0}\}; \\
&&\{p\in \Delta _{n}^{-}\}\quad \Leftrightarrow \quad
\{T^{-1}p,T^{-2}p,\ldots ,T^{-n}p\in \Delta _{0}\}.
\end{eqnarray}%
Sets $\Delta _{n}^{\pm }$ are nested in the following sense:
\begin{eqnarray}
&&\ldots \subset \Delta _{n+1}^{+}\subset \Delta _{n}^{+}\ldots \subset
\Delta _{1}^{+}\subset \Delta _{0} \\
&&\ldots \subset \Delta _{n+1}^{-}\subset \Delta _{n}^{-}\ldots \subset
\Delta _{1}^{-}\subset \Delta _{0}.
\end{eqnarray}%
Now, we define sets
\begin{equation}
\Delta ^{+}=\bigcap_{n=1}^{\infty }\Delta _{n}^{+},\quad \Delta
^{-}=\bigcap_{n=1}^{\infty }\Delta _{n}^{-}.
\end{equation}%
Consider set $\Delta =\Delta ^{+}\cap \Delta ^{-}$. It is is invariant with
respect to the action of the $T$ map. Orbits generated by points from $%
\Delta $ are in one-to-one correspondence with non-collapsing solutions of
Eq. (\ref{eq:stationary}). Therefore, the numerical study of sets $\Delta
_{n}^{\pm }$ allows one to predict and compute bounded solutions of Eq. (\ref%
{eq:stationary}).

There are several restrictions for $Q(x)$ and $P(x)$ for this approach to be
applicable. In Ref. \cite{AlfLeb}, the following statements were proved.

\begin{ntheorem}
Suppose that $Q(x), P(x) \in C^1(\mathbb{R})$ and for each $x \in \mathbb{R}$

\begin{itemize}
\item[a)] there exists ${\widetilde{P}}$ such that $P(x) > 0$, $|P^{\prime
}(x)| \le \widetilde{P}$;

\item[b)] there exist $Q_0, \widetilde{Q}$, such that $Q(x) \ge Q_0$, $%
|Q^{\prime }(x)| \le \widetilde{Q}$;
\end{itemize}

then the solution to the Cauchy problem for Eq.(\ref{eq:stationary}) with
arbitrary initial conditions (\ref{initial}) can be continued onto the whole
real axis $\mathbb{R}$. \label{prop:nonsingular}
\end{ntheorem}

\begin{ntheorem}
Suppose that $\forall {x\in \mathbb{R}}$ conditions $P(x)<0$, $Q(x)<0$
holds, then all solutions of Eq. (\ref{eq:stationary}) are singular, except
the trivial zero solution. \label{prop:singular}
\end{ntheorem}

In particular, this implies that, if $P(x)$ and $Q(x)$ are bounded and
periodic, and $P(x)>0$ for all $x\in \mathbb{R}$, then all solutions of Eq. (%
\ref{eq:stationary}) are non-singular, and the present approach cannot be
applied. In the case of $P(x)<0$, $Q(x)<0$, Eq. (\ref{eq:stationary}) has no
non-singular solutions, except for the zero state, therefore the approach
cannot be used either. However, it follows from Proposition 2 of Ref. \cite%
{AlfLeb} that, if $P(x)$ is a sign-alternating function, the collapsing
behavior is \textit{generic} for solutions of Eq. (\ref{eq:stationary}), and
the application of the approach is reasonable for finding non-collapsing
solutions.

In Ref. \cite{AlfAvr} the case of $P(x)=-1$ in Eq. (\ref{eq:stationary}) was
considered from a more abstract viewpoint. It was shown that if

\begin{itemize}
\item[a)] the $\Delta _{0}$ set consist of a finite number $N$ of connected
components, {\color{black} $\Delta_{0}=\bigcup_{k=1}^{N} D_k$}, and each of the components $%
D_{k}$ is a curvilinear quadrangle, whose boundaries satisfy special
conditions of smoothness and monotonicity;

\item[b)] all the sets $TD_{k}\cap D_{m}$ and $T^{-1}D_{k}\cap D_{m}$, $%
k,m=1,\dots ,N$, are non-empty, and the action of $T$ on curves lying in $%
D_{k}$ preserves the monotony property;

\item[c)] areas of sets $\Delta _{n}^{\pm }$ vanish at $n\rightarrow \infty $%
;
\end{itemize}

then orbits of the Poincar\'{e} map $T$ acting on the $\Delta _{0}$ set are
in one-to-one correspondence with bi-infinite sequences of symbols of some $%
N $-symbol alphabet.

This result can be commented as follows. Let symbols of the alphabet be the
numbers $1,\dots ,N$. Denote the connected components of $\Delta _{0}$ by $%
D_{k}$, $k=1,\ldots ,N$. Then for each non-collapsing solution $u(x)$ there
exist an unique orbit $\{p_{k}\}$, $k=0,\pm 1,\pm 2,\ldots $, $p_{k}\in
\Delta $, and the corresponding unique bi-infinite sequence $\ldots \alpha
_{-1},\alpha _{0},\alpha _{1},\ldots $, $\alpha _{k}\in \{1,\ldots ,N\}$
such that
\begin{eqnarray}
\ldots,~&p_{-1}=T^{-1}p_{0}\in D_{\alpha _{-1}},\quad
p_{0}\in D_{\alpha_{0}},\nonumber\\ 
&p_{1}=Tp_{0}\in D_{\alpha _{1}},\ldots\label{Seq} 
\end{eqnarray}%
On the contrary, for each bi-infinite sequence of numbers $\{1,\ldots ,N\}$
there exists an unique orbit $\{p_{k}\}$, $k=0,\pm 1,\pm 2,\ldots $, $%
p_{k}\in \Delta $, that satisfies condition (\ref{Seq}) and corresponds to an
unique solution $u(x)$. The check of conditions (a),(b) and (c) was carried
out in Ref. \cite{AlfAvr} numerically, using some auxiliary statements.

In what follows below, we apply this approach to Eq. (\ref{eq:stationary})
with $U(x)=0$, i.e., $Q(x)=\omega $, when the linear potential is absent,
and only the pseudopotential is present in Eq. (\ref{eq:stationary}),
induced by the modulation function taken as
\begin{equation}
P(x)=\alpha +\cos (2x),  \label{NP}
\end{equation}%
This is a new setting for which the present method was not elaborated
previously.

\subsection{Numerical results}

\label{NumSteady}

According to what was said above [Eq. (\ref{NP})], we now focus on the
following version of Eq. (\ref{eq:stationary}):
\begin{equation}
u_{xx}+\omega u+(\alpha +\cos 2x)u^{3}=0.  \label{eq:current}
\end{equation}%
Due to Theorem \ref{prop:nonsingular} we impose restriction $\alpha \in
(-1,1)$ in Eq. (\ref{eq:current}) for the approach to be applied, i.e., the
nonlinearity coefficient (\ref{NP}) must be a sign-changing function of $x$.
Another restriction, $\omega <0$, comes from the obvious condition of the
soliton localization, given by Eq. (\ref{Local}).

\begin{figure}%[tbp]
\includegraphics[scale=0.45]{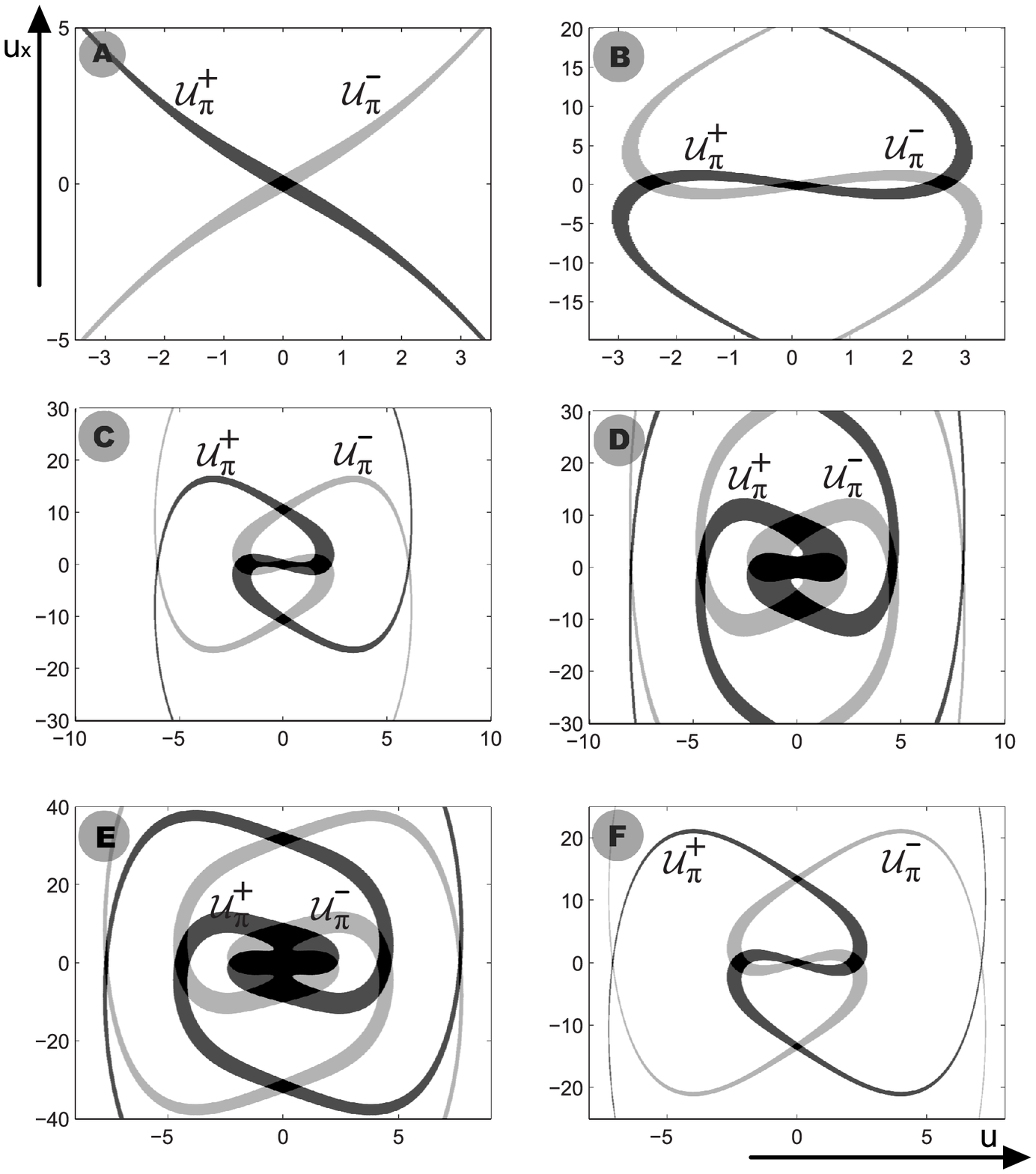}
\caption{$\mathcal{U}_{\protect\pi }^{+}$ (dark grey color), $\mathcal{U}_{%
\protect\pi }^{-}$ (light grey color) and their intersection $\Delta _{0}$ {\color{black} (black color)}
in the model based on Eq. (\protect\ref{eq:current}), at different values of
parameters $\protect\omega $ and $\protect\alpha $: A) $\protect\omega =-1$,
$\protect\alpha =-1.1$; B) $\protect\omega =-1$, $\protect\alpha =-0.3$; C) $%
\protect\omega =-1$, $\protect\alpha =0.15$; D) $\protect\omega =-1$, $%
\protect\alpha =0.5$; E) $\protect\omega =-0.7$, $\protect\alpha =0.55$; F) $%
\protect\omega =-1.5$, $\protect\alpha =0$.}
\label{pic:spirals}
\end{figure}

\textit{Sets $\mathcal{U}_{\pi }^{\pm }$.} The set $\mathcal{U}_{\pi }^{+}$
was found by scanning the plane $(u,u^{\prime })$ of initial data by means
of the following procedure. The Cauchy problem for Eq. (\ref{eq:current})
was solved numerically, taking as initial conditions $u(0)=n\Delta u$, $u_x(0)=m\Delta u^{\prime }$, $m,n=-L,\ldots ,L$ where spacings $\Delta u$ and $\Delta u^{\prime }$ are 
small enough (typical values were $\Delta u=\Delta
u^{\prime }=0.01$). If the absolute value of the solution of the Cauchy
problem exceeds, in interval $[0;\pi ]$, some sufficiently large value $%
u_{\infty }$, it is assumed that the collapse occurs. The corresponding
point is marked {\color{black} white}, otherwise, it is {\color{black} grey}. The computations were
actually performed for $u_{\infty }=10^{5}$ and further checked for $%
u_{\infty }=10^{7}$, the results obtained for both cases agreeing very well.
Since Eq.(\ref{eq:current}) is invariant with respect to inversion $x\to-x$,
the set $\mathcal{U}_{\pi }^{-}$ is the reflection of $\mathcal{U}_{\pi }^{+}
$ with respect to the $u$-axis. The numerical results allow us to conjecture
that, for $\alpha \in (-1;1)$, $\mathcal{U}_{\pi }^{\pm }$ \textit{are
unbounded spirals with infinite number of rotations around the origin}, see
Fig. \ref{pic:spirals}.

\textit{Set $\Delta _{0}$.} Some examples of set $\Delta _{0}$ are displayed
in Fig.~\ref{pic:spirals}. Panel (A) of Fig. \ref{pic:spirals} corresponds
to the case of $\omega =-1$, $\alpha =-1.1$, when $\Delta _{0}$ consists of
only one connected component situated in the origin. This fact agrees with
Theorem \ref{prop:singular}. If $\alpha \in (-1;1)$, then, presumably, $%
\Delta _{0}$ is unbounded and consists of an infinite number of connected
components that are situated along the $u$ and $u^{\prime }$ axes [panels
(B)-(F) of Fig. \ref{pic:spirals}]. The connected components can be
enumerated by symbols $\{A_{k}\},k=\pm 1,\pm 2,\dots $ (the components along
$u$ axis) and $\{B_{k}\},k=\pm 1,\pm 2,\dots $ (the components along $%
u^{\prime }$ axis). The central connected component is denoted $O$.
The basic assumption for the applicability of the coding approach is that
the the connected components are curvilinear quadrangles with opposite sides
lying on the boundaries of $\mathcal{U}_{\pi }^{+}$ and $\mathcal{U}_{\pi
}^{-}$. Due to geometric properties of the spirals, it is quite natural to
assume that all connected components $\{A_{k}\},\{B_{k}\},k=\pm 1,\pm
2,\dots $ satisfy this condition. However, central connected component $O$
may be such a curvilinear quadrangle (cases A, B, F in Fig. \ref{pic:spirals}%
), or may be not (cases C, D, E in Fig. \ref{pic:spirals}), depending on
values of $\omega $ and $\alpha $.

\textit{Coding.} Assume that the parameters $\omega $ and $\alpha $ are such
that all connected components in $\Delta _{0}$ are curvilinear quadrangles.
Then, our numerical study indicates that $T^{-1}A_{k}$, $T^{-1}B_{k}$, $%
k=1,2,\ldots $, and $T^{-1}O$ are infinite curvilinear strips situated
inside $\mathcal{U}_{\pi }^{+}$ and crossing all the connected components.
Similarly, $TA_{k}$, $TB_{k}$, $k=1,2,\ldots $, and $TO$ are also
curvilinear strips situated inside the $\mathcal{U}_{\pi }^{-}$ set, that
also cross all the connected components. $T$-pre-images of the sets
\begin{eqnarray*}
&&T^{-1}Z\cap A_{l},\quad T^{-1}Z\cap B_{l},\quad T^{-1}Z\cap O,\quad l=\pm
1,\pm 2,\ldots , \\
&&Z\in \{O,A_{k},B_{k},k=\pm 1,\pm 2,\ldots \},
\end{eqnarray*}%
are infinite curvilinear strips belonging to $T^{-1}Z$. Similar statement
are also valid for $T$-images of $TZ\cap A_{l}$, $TZ\cap B_{l}$, $TZ\cap O$ $%
l=\pm 1,\pm 2,\ldots $ which are placed inside $TZ$, with $Z\in
\{O,A_{k},B_{k},k=\pm 1,\pm 2,\ldots \}$. Therefore the situation is similar
to one considered in Ref.\cite{AlfAvr} and we conjecture that the dynamics
of $T$ is similar to dynamics of the Poincar\'{e} map from Ref.\cite{AlfAvr}%
, and that \textit{there is one-to-one correspondence between all
nonsingular solutions of Eq.(\ref{eq:current}) and bi-infinite sequences $%
\{\ldots Z_{-1},Z_{0},Z_{1},\ldots \}$ based on the infinite alphabet of
symbols $Z_{m}\in \{O,A_{k},B_{k},k=\pm 1,\pm 2,\ldots \}$}. The orbit
corresponding to code $\{\ldots ,Z_{-1},Z_{0},Z_{1},\ldots \}$ visits
successively connected components $Z_{m}$, $m=\ldots ,-1,0,1,\ldots $. Note
that the orbit corresponding to the soliton solution starts and ends in the
central connected component, therefore it has the code of the form $\{\ldots
,O,O,Z_{1},Z_{2},\ldots ,Z_{N},O,O,\ldots \}$ where symbols $Z_{1}$ and $%
Z_{N}$ are different from $O$.

\begin{figure}%[tbp]
\includegraphics[scale=0.4]{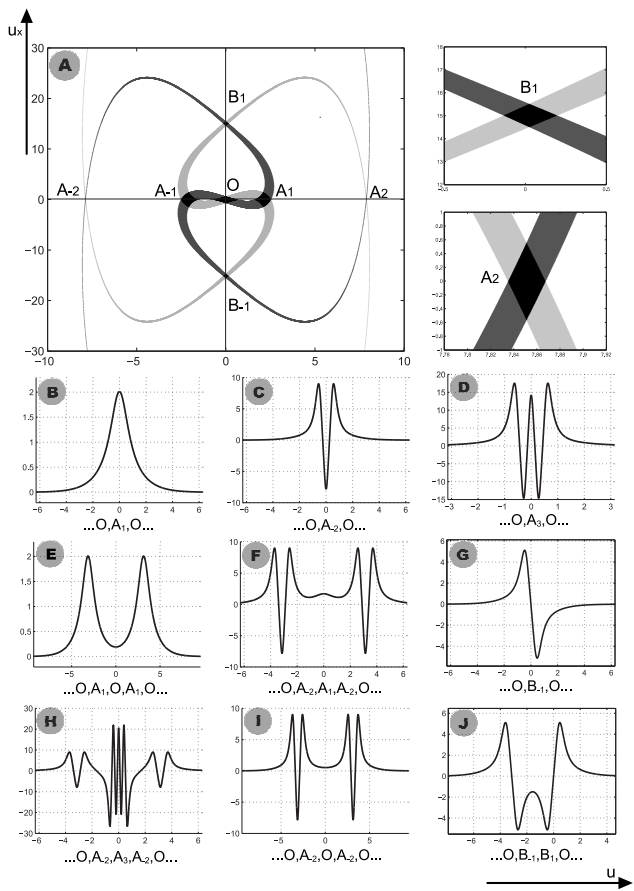}
\caption{Numerically found solutions of Eq. (\protect\ref{eq:current}) and
their codes for parameters $\protect\omega =-1$, $\protect\alpha =-0.1$; A) $%
\mathcal{U}_{\protect\pi }^{\pm }$ sets; {\color{black} B)-J) the profiles of solitons together with their codes}
% $\{\dots OA_{1}O\dots \}$; C) $%
%\{\dots OA_{-2}O\dots \}$; D) $\{\dots OA_{3}O\dots \}$; E) $\{\dots
%OA_{1}OA_{1}O\dots \}$; F) $\{\dots OA_{-2}A_{1}A_{-2}O\dots \}$; G) $%
%\{\dots OB_{-1}O\dots \}$; H) $\{\dots OA_{-2}A_{3}A_{-2}O\dots \}$; I) $%
%\{\dots OA_{-2}OA_{-2}O\dots \}$; J) $\{\dots OB_{-1}B_{1}O\dots \}$. 
}
\label{pic:coding}
\end{figure}

\textit{Solitons.} Regardless of whether the coding conjecture is true or
false generically, it might be used for the prediction of possible shapes of
nonlinear modes. Specifically, the location of the connected components in
the plane of $(u,u^{\prime })$, and the order in which the orbit visits
them, yields comprehensive information about the nonlinear mode. In the
present model, the predicted nonlinear modes were found numerically in
\textit{all} the cases considered. Some of soliton solutions of Eq.(\ref%
{eq:current}) and their codes are shown in Fig. \ref{pic:coding} for $\omega
=-1$, $\alpha =-0.1$. The soliton in panel (B) is the FS, cf. Ref. \cite%
{Malomed}, with code {\color{black} $\{\dots, O,A_{1},O,\dots \}$, or $\{\dots,O,A_{-1},O,\dots \}$}%
, which is its symmetric counterpart. Another particular solution, 
shown in panel G, represents the above-mentioned DSs (\textit{dipole solitons}), 
which are essentially confined to
a single cell of the nonlinear lattice. This solution corresponds
to code {\color{black} $\{\dots, O,B_{-1},O,\dots, \}$}, and its symmetric counterpart is {\color{black} $\{\dots,O,B_{1},O,\dots \}$}. The DSs are similar to the
(mostly unstable) SFSs reported in Refs. \cite{Thaw}-\cite{China} in
models with the linear lattice potential, as both soliton species feature 
the antisymmetric profile squeezed into a single cell of the underlying lattice
(the linear one, in the case of the SFSs, and the nonlinear lattice, 
as concerns the DSs). The area of the localization of
the soliton corresponding to code $\{\ldots ,O,Z_{1},Z_{2},\ldots
,Z_{N},O,\ldots \}$, where the symbols $Z_{1}$ and $Z_{N}$ are different
from $O$, is $N\pi $, i.e., it extends over $N$ periods of the underlying
nonlinear lattice. In particular, the solitons with codes $\{\ldots
,O,O,Z,O,O,\ldots \}$, $Z\neq O$ (named \textit{elementary solitons }in what
follows below), are localized, essentially, in one period of the lattice.

\section{The linear-stability analysis}

\label{LinStab}

%\subsection{Spectral problem}

As said above, stability is a critically important issue for solitons
supported by lattice potentials. Here, we address the
stability of solitons produced by Eqs. (\ref{UP}), (\ref{eq:current}). It
has been shown in Sect. \ref{StatMod} that there exist a great variety of
shapes of such modes. Thus, adopting the nonlinear lattice as given by Eq. (%
\ref{NP}), we aim to study the linear stability of solitons generated by
equation
\begin{equation}
i\Psi _{t}+\Psi _{xx}+(\alpha +\cos 2x)|\Psi |^{2}\Psi =0
\label{eq:stability}
\end{equation}

Following the well-established approach, (see, e.g., Ref. \cite{JYang}), we
consider small perturbations around a stationary solution $\Psi
_{0}(x,t)=u(x)e^{-i\omega t}$ in the form of
\begin{equation}
\Psi (t,x)=\left[ u(x)+\widetilde{U}(t,x)\right] e^{-i\omega t},~\left\vert
\widetilde{U}(t,x)\right\vert \ll 1,
\end{equation}%
where $u(x)$ is a localized solution of Eq. (\ref{eq:current}), and the
perturbation satisfies the linear equation
\begin{equation}
i\widetilde{U}_{t}+\widetilde{U}_{xx}+\omega \widetilde{U}+(\alpha +\cos
2x)u^{2}(2\widetilde{U}+\widetilde{U}^{\ast })=0,  \label{linear}
\end{equation}%
where asterisk means complex conjugate. Seeking solutions to Eq. (\ref%
{linear}) as
\begin{equation}
\widetilde{U}(t,x)=(v(x)+w(x))e^{\lambda t}+(v^{\ast }(x)-w^{\ast
}(x))e^{\lambda ^{\ast }t},
\end{equation}%
we arrive at the eigenvalue problem
\begin{equation}
LY=\lambda Y,  \label{eq:eigproblem}
\end{equation}%
\begin{equation}
L=i%
\begin{pmatrix}
0 & L_{0} \\
L_{1} & 0%
\end{pmatrix}%
,~~~Y=%
\begin{pmatrix}
v \\
w%
\end{pmatrix}%
,  \label{Y}
\end{equation}%
where
\begin{eqnarray*}
&&L_{0}=\partial _{xx}+G_{0}(x),\quad G_{0}(x)=\omega +(\alpha +\cos
2x)u^{2},\\  %\notag \\
&&L_{1}=\partial _{xx}+G_{1}(x),\quad G_{1}(x)=\omega +3(\alpha +\cos
2x)u^{2}. % \label{G}
\end{eqnarray*}%
The soliton is linearly unstable if the spectrum produced by Eq. (\ref%
{eq:eigproblem}) contains at least one eigenvalue $\lambda $ with a non-zero
real part, $\Re (\lambda )>0$. Otherwise, the solitons are linearly stable.

Equation (\ref{eq:eigproblem}) generates the spectrum consisting of
continuous and discrete parts. It is easy to show that the continuous
spectrum is represented by two rays, $[-i\omega ;+i\infty )$ and $(-i\infty
;i\omega ]$, if $\omega <0$, and by the whole imaginary axis, if $\omega >0$%
. The discrete spectrum includes zero eigenvalue $\lambda =0$. Other
eigenvalues of the discrete spectrum appear in quadruples, since if $\lambda
$ is an eigenvalue then $-\lambda $, $\lambda ^{\ast }$ and $-\lambda ^{\ast
}$ are eigenvalues too.

To find discrete eigenvalues numerically, the Fourier Collocation Method (FCM)
\cite{JYang} was used. This method is very efficient to find 
{\it exponential instabilities}, that appears due to real eigenvalues. However it
is known that it can miss the situations of weak {\it  oscillatory instabilities}
caused by quartets of complex eigenvalues with small real parts (see e.g. \cite{Egor})
where more sophisticated methods, such as Evans function method, \cite{PelKiv}, must be applied.  With the help of FCM, a great number of stationary solutions of Eq.(\ref{eq:stability}%
), represented by different codes, were analyzed. Due the infinite number of
essentially different solutions, it is not possible to perform a
comprehensive stability analysis of all localized solutions, even of all
elementary solitons. However, we observed that a majority of the solitons
are linearly unstable, thus being physically irrelevant solutions. \emph{%
Stable solitons} can be categorized as follows:

a) Among the elementary solitons, it was found that FS and DS are \textit{%
linearly stable}, under some restrictions on $\omega $ and $\alpha $. Other
elementary solitons were found to be unstable. Note that
FSs are considered as stable solutions in models with linear lattice potentials, see Ref. \cite{Malomed}
and references therein, while the SFSs are chiefly unstable in that case, having a small stability 
region \cite{China} (strictly speaking, FSs in models with linear lattice potentials may also feature a very weak oscillatory instability, having at the same time great lifetime, see \cite{Egor}). Therefore, \emph{stable} DSs supported by the nonlinear
pseudopotential, whose shape is very similar to that of the \emph{chiefly unstable} SFSs in
the systems with linear lattice potentials, deserve a detailed consideration, which is given in Sect. %
\ref{SFS}. It includes not only numerical results, but also analytical ones
based on VA.

b) There are stable bound states of FSs -- for instance, with codes {\color{black} $\{\dots
,O,A_{1},A_{-1},A_{1},O,\dots \}$, $\{\dots ,O,A_{1},O,A_{-1},O,\dots \}$}.
However, other bound states of these modes may be unstable.

\begin{figure*}[tbp]
\subfloat[][$\{\dots,O,A_1,O,\dots\}$]{\includegraphics[width=0.5%
\textwidth]{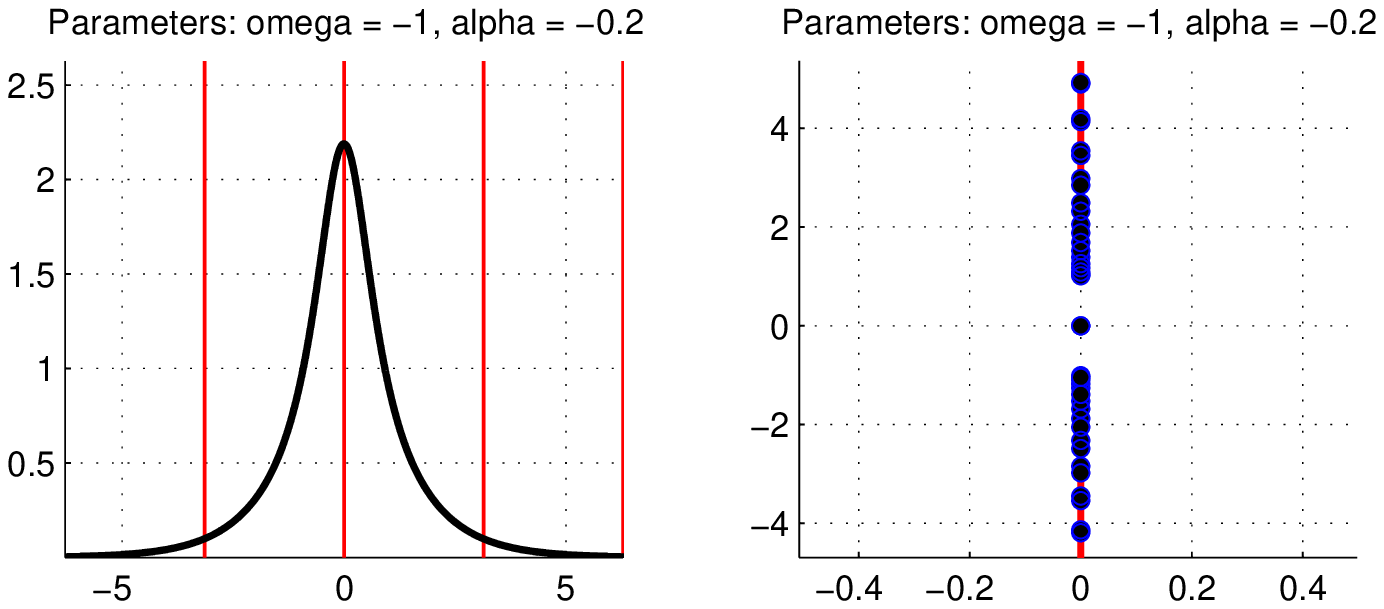}} \subfloat[][$\{\dots,O,B_{-1},O,\dots\}$]{%
\includegraphics[width=0.5\textwidth]{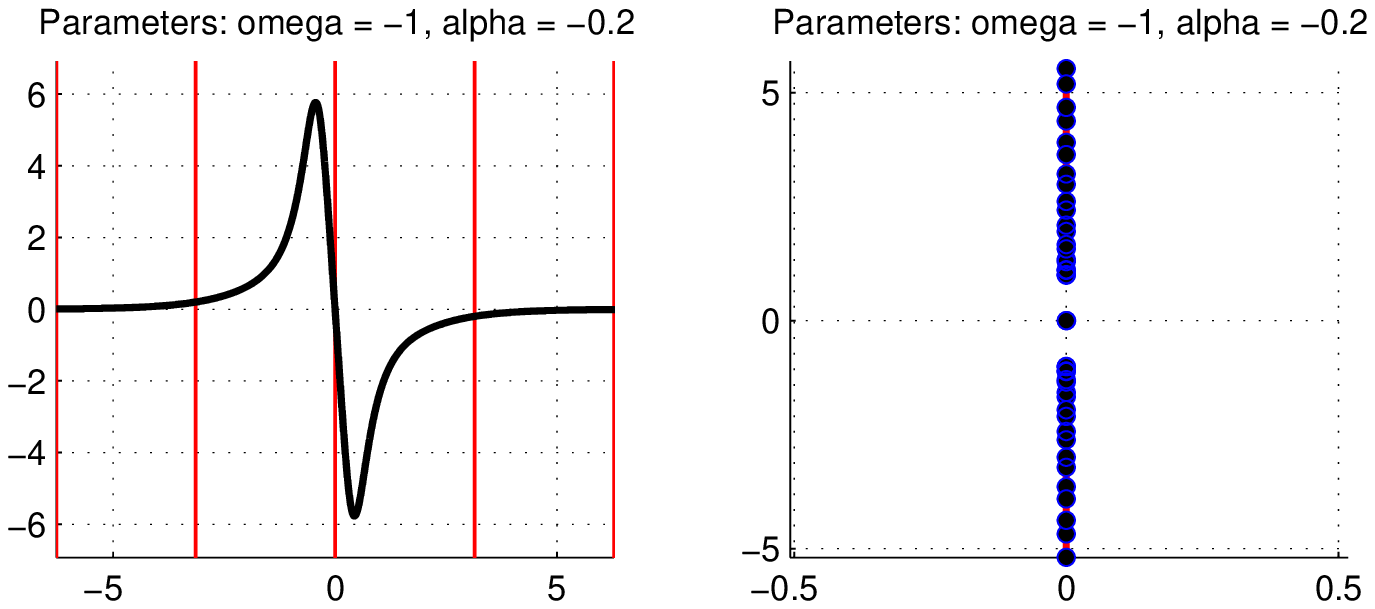}} \vfill
\subfloat[][$\{\dots,O,A_1,A_{-1},A_1,O,\dots\}$]{\includegraphics[width=0.5%
\textwidth]{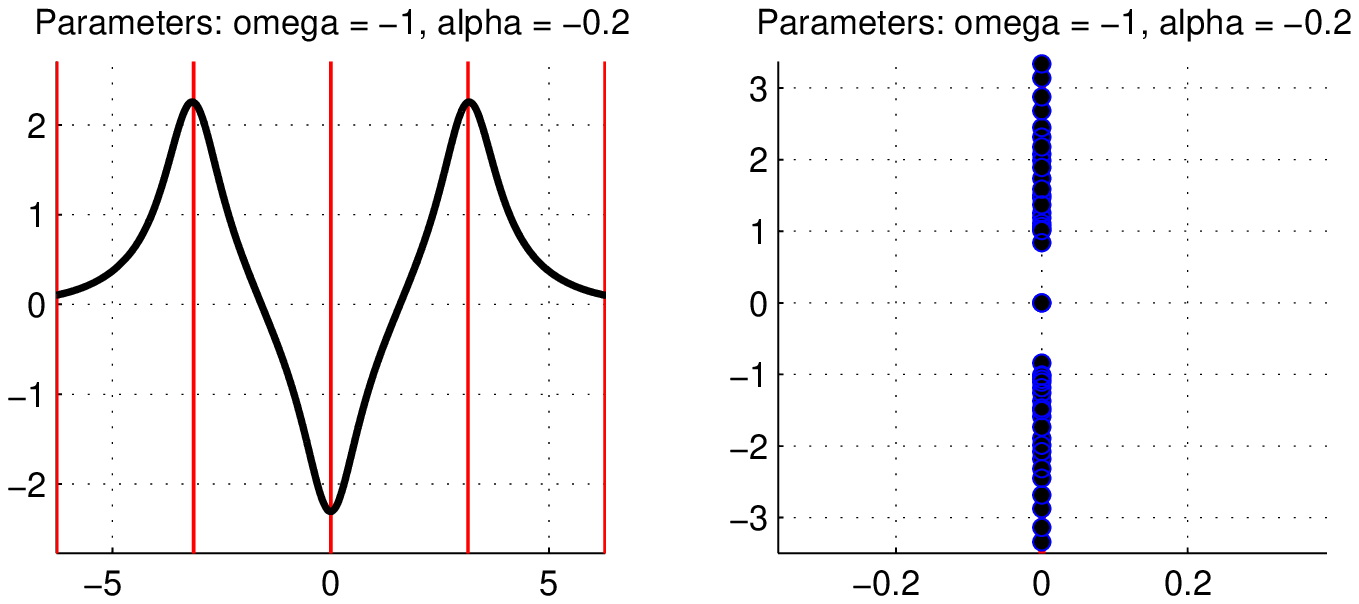}} \subfloat[][$\{\dots,O,A_1,A_1,A_1,O,\dots\}$]{%
\includegraphics[width=0.5\textwidth]{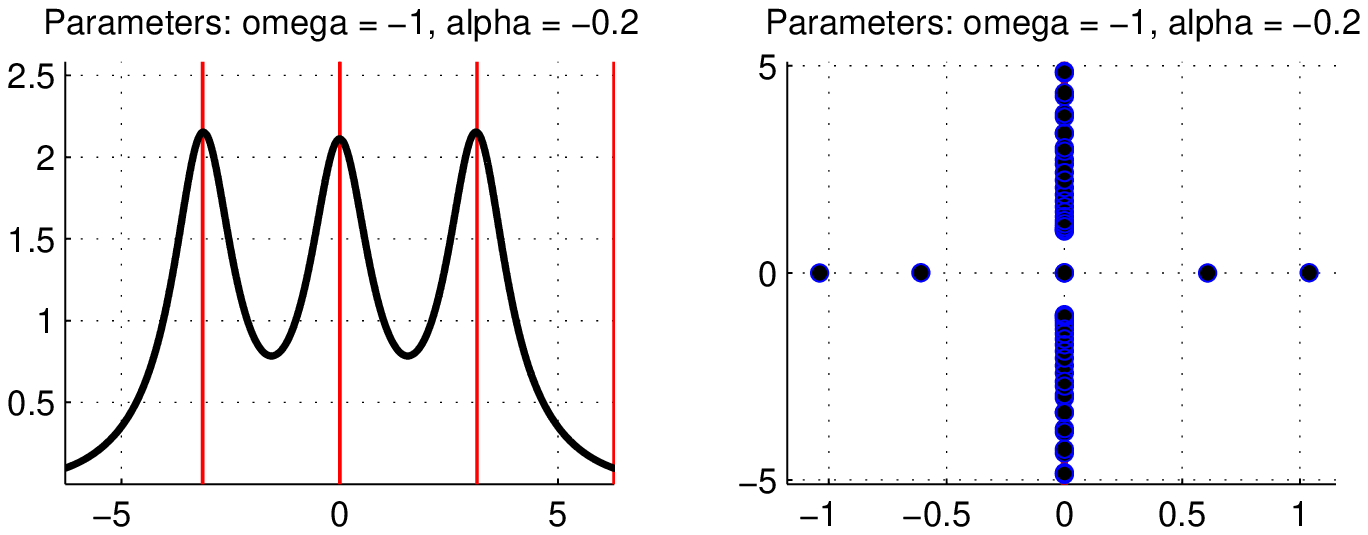}} \vfill
\subfloat[][$\{\dots,O,A_1,A_1,O,\dots\}$]{\includegraphics[width=0.5%
\textwidth]{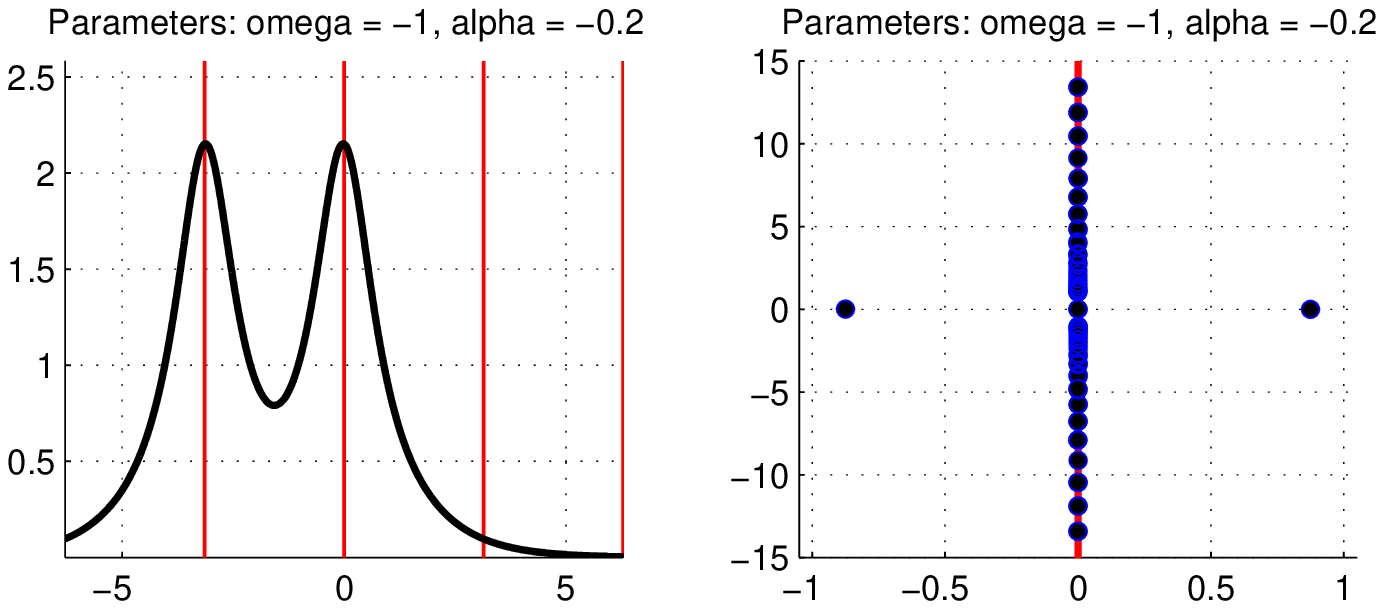}} \subfloat[][$\{\dots,O,A_1,O,A_{-1},O,\dots\}$]{%
\includegraphics[width=0.5\textwidth]{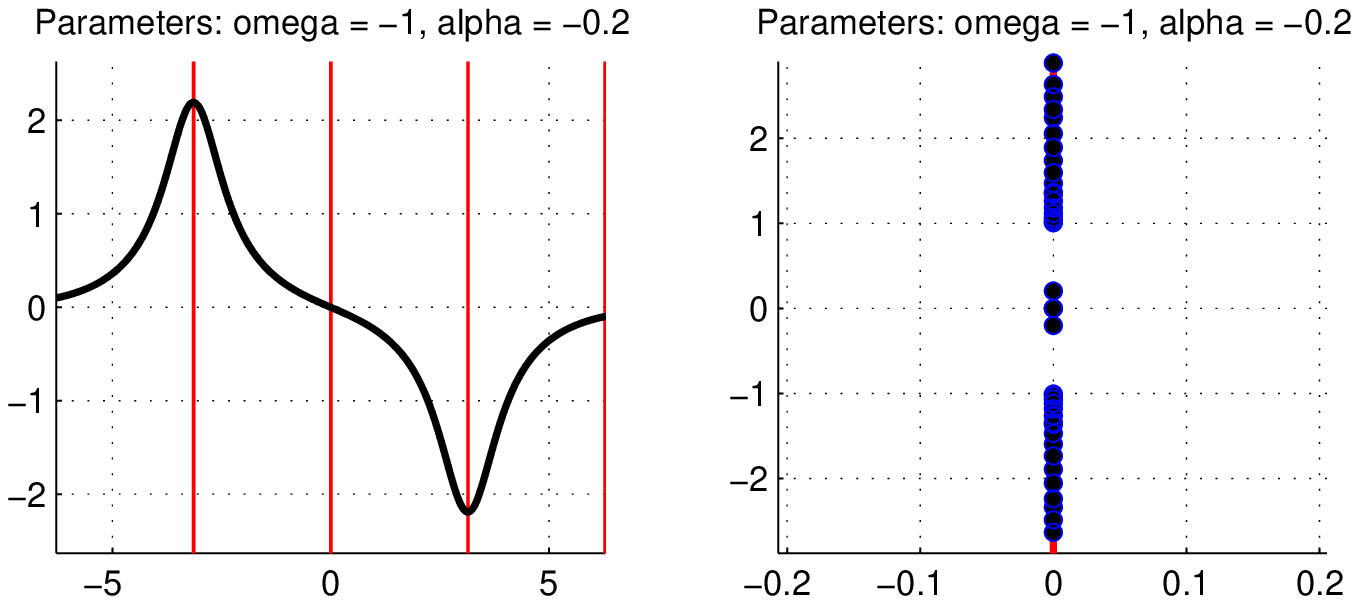}} \vfill
\subfloat[][$\{\dots,O,A_1,B_1,A_{-1},O,\dots\}$]{\includegraphics[width=0.5%
\textwidth]{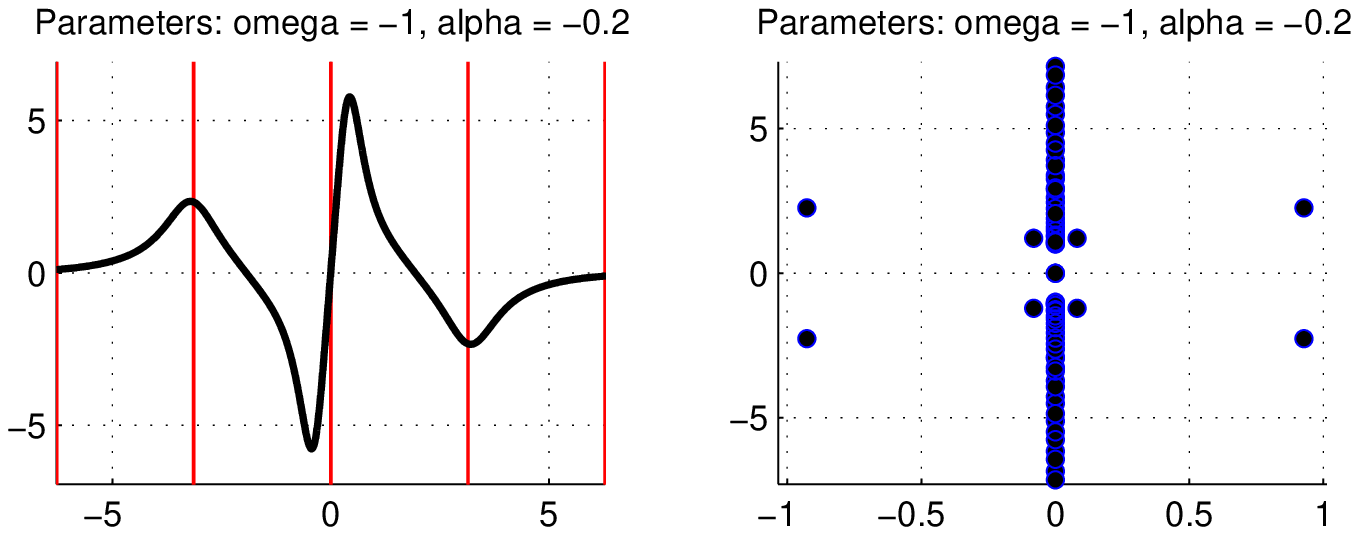}} \subfloat[][$\{\dots,O,A_1,B_{-1},A_{-1},O,%
\dots\}$]{\includegraphics[width=0.5\textwidth]{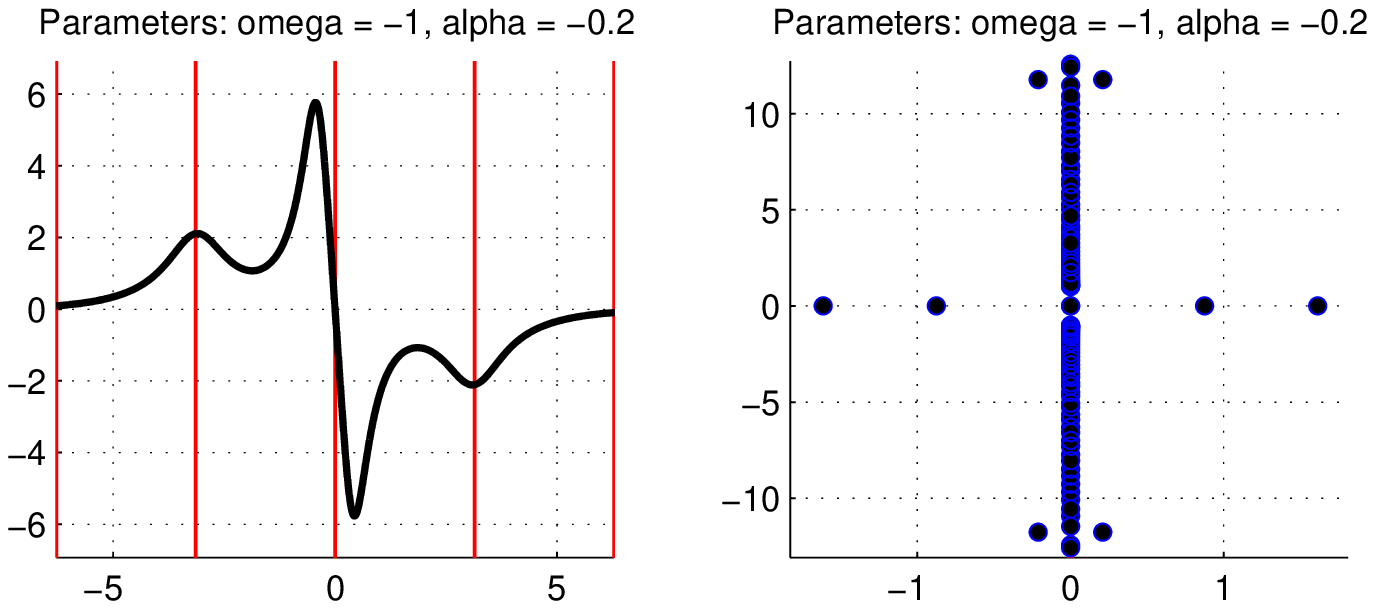}}
\caption{Localized solutions, their codes, and linear-stability spectra for $%
\protect\omega =-1$, $\protect\alpha =-0.2$.}
\label{pic:stability}
\end{figure*}

Stability spectra for some solitons and their bound states are shown in Fig.%
\ref{pic:stability}. These example adequately represent the generic
situation.
%\onecolumngrid
%\widetext{
%}
%\twocolumngrid

\section{Dipole solitons (DSs)}

\label{SFS}

\subsection{The variational approximation}

\label{SFS_VA}

Some general features of soliton solutions of Eq. (\ref{eq:current}) can be
obtained by means of the VA, using the fact that Eq. (\ref{eq:current}) for
the stationary states can be derived from Lagrangian
\begin{equation}
L=\int_{-\infty }^{+\infty }\left\{ \frac{1}{2}\left( u^{\prime }\right)
^{2}-\frac{1}{2}\omega u^{2}-\frac{1}{4}\left[ \alpha +\cos \left( 2x\right) %
\right] u^{4}\right\}~dx .  \label{eq:lagrangian}
\end{equation}%
In Ref. \cite{Malomed}, VA was successfully applied for analysis of FS. In
that study, the soliton was assumed to be bell-shaped, and the following
ansatz was used
\begin{equation}
u(x)=A\exp \left( -\frac{x^{2}}{2W^{2}}\right) ,
\end{equation}%
The VA had yielded correct predictions for the existence of the minimal norm
\begin{equation}
N= \int_{-\infty }^{+\infty }u^{2}(x)dx=\sqrt{\pi }A^{2}W.
\end{equation}%
for the FS, and the existence of an amplitude threshold for stable solitons.

A similar analysis for the DS may be based on the simplest spatially odd
ansatz:
\begin{equation}
u(x)=Ax\exp \left( -\frac{x^{2}}{2W^{2}}\right) ,  \label{eq:ansatz}
\end{equation}%
The maximum value of $u(x)$, which is $\sqrt{e}AW$, is situated at $x_{\max
}=W$, therefore $W$ may be regarded as the half-width of the DS. Norm $N$
of ansatz (\ref{eq:ansatz}) is
\begin{equation}
N=\frac{\sqrt{\pi }}{2}A^{2}W^{3}.  \label{eq:norm}
\end{equation}%
Equation (\ref{eq:norm}) makes it possible to eliminate the amplitude $A$ in
favor of the norm:%
\begin{equation}
A^{2}=\frac{2}{\sqrt{\pi }}\frac{N}{W^{3}}.  \label{eq:amplitude}
\end{equation}

The substitution of ansatz (\ref{eq:ansatz}) into Lagrangian (\ref%
{eq:lagrangian}) and calculation of the integrals yields the following
effective Lagrangian:
\begin{eqnarray}
&&L_{\mathrm{eff}}=-\frac{\omega }{2}N+\frac{3N}{4W^{2}}-\frac{3\alpha N^{2}}{%
16\sqrt{2\pi }W}-\nonumber\\[2mm]
&&-\frac{N^{2}e^{-W^{2}/2}}{16\sqrt{2\pi }W}\left(
3-6W^{2}+W^{4}\right) ,  \label{eq:leff}
\end{eqnarray}%
where Eq. (\ref{eq:amplitude}) was used to eliminate $A^{2}$. The
Euler-Lagrange (variational) equations following from the effective
Lagrangian are
\begin{eqnarray}
\partial L_{\mathrm{eff}}/\partial W &=&0,  \label{eq:dw} \\
\partial L_{\mathrm{eff}}/\partial N &=&0,  \label{eq:dn}
\end{eqnarray}%
with $W$ and $N$ treated as free variational parameters for given $\omega $.$%
\allowbreak $

\begin{figure}[tbp]
\center{\includegraphics[width=0.45\textwidth]{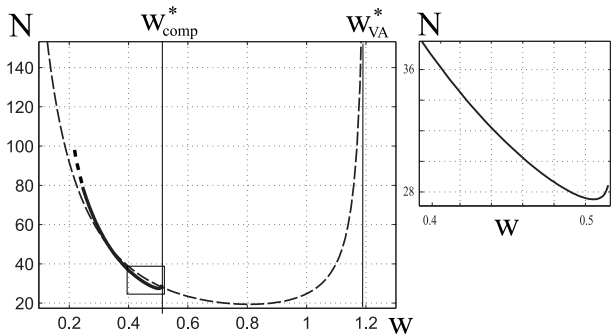}}
\caption{\textcolor{black}{Left panel: the relation between the norm and width of the DS, as predicted by the variational approximation,  (thin dashed line, $\alpha=0$). Bold line shows the same relation for numerically computed DS. Right panel: the magnification of the bold line in small rectangle in the left panel}}
\label{pic:N(W)}
\end{figure}

Hereafter, we consider the case $\alpha =0$ in more detail. Equation (\ref%
{eq:dw}) implies the following relation between $N$ and $W$:
\begin{equation}
N=\frac{48\sqrt{\pi /2}\exp \left( W^{2}/2\right) }{W\left(
3+9W^{2}-9W^{4}+W^{6}\right) }.  \label{eq:N(W)}
\end{equation}%
This relation is plotted in Fig. \ref{pic:N(W)},\textcolor{black}{ (left panel, thin dashed line),} where it attains a minimum
value,
\begin{equation}
N_{\min }^{(\mathrm{VA})}\approx 19.41  \label{min-VA}
\end{equation}%
at $W=W_{0}\approx 0.806$.

An essential feature of the dependence is that it predicts the existence of
a minimum norm necessary for the DS to exist. Furthermore, it follows from
Eq.(\ref{eq:N(W)}) that the range of the variation of $W$ predicted by the
VA is \emph{finite}:
\textcolor{black}{\begin{equation}
0<W<W^*_{VA}\approx 1.21.  \label{eq:wmax}
\end{equation}
}

The second variational equation, Eq.(\ref{eq:dn}), yields, after additional
algebraic manipulations, a monotonic dependence $\omega $ on $W$:
\begin{equation}
\omega =\frac{3}{2}\cdot\frac{-9+33W^{2}-13W^{4}+W^{6}}{W^{2}\left(
3+9W^{2}-9W^{4}+W^{6}\right) }.  \label{eq:omega}
\end{equation}%
It may be combined with Eq. (\ref{eq:N(W)}) to apply the VK stability
criterion \cite{Vakh}, $dN/d\omega \equiv \left( d\omega /dW\right)
^{-1}dN/dW<0$. Because it follows from Eq. (\ref{eq:omega}) that $d\omega
/dW $ is always positive, the VK criterion predicts that stable is the left
branch in Fig. \ref{pic:N(W)}, with $dN/dW<0$, which corresponds to interval%
\begin{equation}
0<W<W_{0}\approx 0.806,  \label{eq:stab}
\end{equation}%
while the right branch, with $dN/dW>0$, i.e., $W>W_{0}$ is unstable.

Note that Eq. (\ref{eq:omega}) is compatible with the above-mentioned
localization condition, $\omega <0$, at $0<W<0.556$, while the fact that the
VA predicts $\omega >0$ at $W>0.556$ is a manifestation of its inaccuracy.
It is worthy to note that the predicted stability region tends to have $%
\omega <0$, i.e., the stability is predicted in the region where the VA is
more accurate.

To summarize, the predictions of VA are:

\noindent (i) the existence of the minimal norm of the DS;

\noindent (ii) the existence of its maximum width;

\noindent (iii) the existence of the maximum width of DSs to be stable.

\noindent In what follows below we show that these predictions qualitatively
agree with results of numerical computation. The application of the VA to
more complex solitons is much more cumbersome and is not presented here.

\subsection{Numerical results for stationary dipole solitons}

\label{The dipole-soliton_profile}

The numerical computation of DS profiles was carried out by dint of the
shooting method. The results can be summarized as follows.

\begin{figure}[tbp]
\centerline{\includegraphics[scale=0.8]{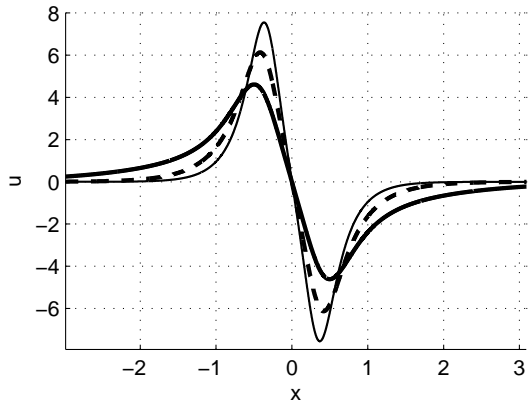}}
\caption{Numerically found profiles of the dipole solitons for $%
\protect\omega =-15$ (thin line), $\protect\omega =-7$ (dash line), and $%
\protect\omega =-1$ (thick line), with $\protect\alpha =0$ in Eq. (\protect
\ref{eq:current}).}
\label{pic:Profiles}
\end{figure}

(1) The DS family of may be parameterized by $\omega $ or by $W$, which is
here defined as the distance of maxima of the wave field from the central
point. The amplitude and norm of the DS grow as the soliton shrinks (i.e.,
when $W$ tends to zero), and in this limit $\omega $ tends to $-\infty $.
Examples of DS profiles for $\alpha =0$ and $\omega =-15$ (thin line), $%
\omega =-7$ (dash line), and $\omega =-1$ (thick line) are depicted in Fig. %
\ref{pic:Profiles}. \textcolor{black}{The dependence of norm $N$ on $W$ is also shown in Fig. \ref{pic:N(W)} (bold line in the left panel and the right panel). It is seen in Fig. \ref{pic:N(W)} that this dependence agrees well with VA results at the interval left to $W_{\rm comp}^*$, the
maximum width of DS. Also it follows from 
Fig. \ref{pic:N(W)} that there is a minimum norm $N_{\min }$ necessary for the existence of the DS, hence the above-mentioned prediction (i) of the VA holds.}

%\begin{figure}[tbp]
%\centerline{\includegraphics[scale=0.8]{Figure05.eps}}
%\caption{The numerically found dependence of $N$ on $W$, the half-width of
%the dipole soliton, at $\protect\alpha =0$.}
%\label{pic:N_W_num}
%\end{figure}

\begin{figure}[tbp]
\centerline{\includegraphics[scale=0.45]{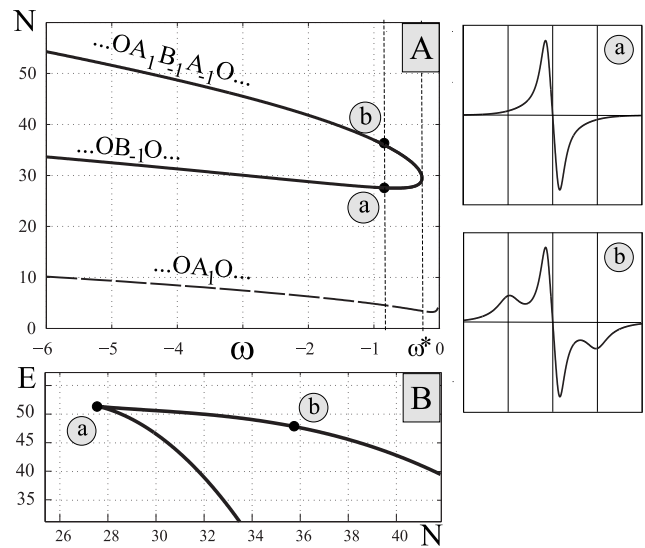}}
\caption{\textcolor{black}{A: The bifurcation diagram for solitons in Eq. (\protect\ref%
{eq:current}) with $\protect\alpha =0$: the family of single-cell dipole
solitons corresponding to code $\{\ldots, O,B_{\pm 1},O,\ldots \}$ coalesces at $%
\protect\omega =\protect\omega ^{\ast }$ with family  $\{\dots, O,A_{\mp
1},B_{\pm 1},A_{\pm 1},O,\dots \}$.  The bottom branch  (dashed line)  represents fundamental solitons, showing that, on the contrary to the 
SFSs in models with linear lattice potentials, the norm of the dipole solitons
is higher than the norm of the fundamental solitons at the same value of $\omega$.
B: Dependence of the energy $E$ on $N$ for the dipole-soliton branch. Two profiles of solitons coexisting at $\protect\omega = -0.8$ are displayed in the right panels (a) and (b), the corresponding points are marked in panels A and B
}}
\label{pic:Bifur}
\end{figure}

(2) The DS exists for $\omega <\omega ^{\ast }$. At {\color{black} $\omega =\omega ^{\ast
} \approx 0.265$ }, the DS family, coded by $\{\ldots, O,B_{\pm 1},O,\ldots, \}$, undergoes a
saddle-node bifurcation and annihilates with the family coded by {\color{black} $\{\dots,
O,A_{\mp 1},B_{\pm 1},A_{\pm 1},O,\dots \}$} (see Fig.~\ref{pic:Bifur}). This
implies that width $W$ of the DSs is bounded from above, hence prediction (ii)
of VA, concerning the existence of the maximum width of the DS, holds too.
\textcolor{black}{However the estimation of VA for the greatest width of the dipol soliton, $W_{VA}^*$,  is quite rough when compared with computed value $W_{\rm comp}^*$, see Fig. \ref{pic:N(W)}.}

Note that the panel A in Fig. \ref{pic:Bifur} also demonstrates that, 
although the single-cell DS is very similar, in its shape, to the SFS in
systems with linear lattice potentials, the DS in the present model is not
subfundamental, as its norm is \emph{higher} than that of the FS existing 
at the same $\omega$. \textcolor{black}{The panel B of  Fig. \ref{pic:Bifur}
presents the dependence of energy $E$ versus the norm $N$. It follows
from  Fig. \ref{pic:Bifur} that the energy for the branch coded by $\{\dots,
O,A_{\mp 1},B_{\pm 1},A_{\pm 1},O,\dots \}$ is greater than the
energy of the DS branch.}

Thus, the predictions of the VA qualitatively agree with the numerical
results, although the accuracy of the VA is rather low, as ansatz (\ref%
{eq:ansatz}) is not accurate enough. For instance, the VA-predicted minimum
norm, given by Eq. (\ref{min-VA}), is smaller than the respective numerical
value,
\begin{equation}
N_{\min }^{\mathrm{(num)}}\approx 27.5,  \label{min-num}
\end{equation}
by $\approx 30\%$. The ansatz may be improved by adding more terms to it,
but then the VA becomes too cumbersome.

\subsection{Evolution of dipole solitons}

\label{SFS_evolution}To check the above-mentioned prediction (iii) of the VA
concerning the stability of the DSs, we have performed simulations of the
evolution of these solitons in the framework of Eq. (\ref{UP}), with $U(x)=0$
and $P(x)$ corresponding to Eq. (\ref{eq:current}). The simulations were run
by means of the Trofimov-Peskov finite-difference numerical scheme \cite%
{Trofimov}. The
scheme is implicit, its realization implying iterations for the calculation
of values in each temporal layer, but it allows running computation with
larger temporal steps. In order to reveal instability (if it is), the
soliton profile was perturbed in initial moment with small spatial
perturbation. A finite spatial domain $[-4\pi, 4\pi]$ was used, with
reflection of radiation from boundaries eliminated by means of absorbing
boundary conditions.

%\onecolumngrid

\begin{figure*}[tbp]
\subfloat[][$\omega =
-0.3$]{\includegraphics[width=0.5\textwidth]{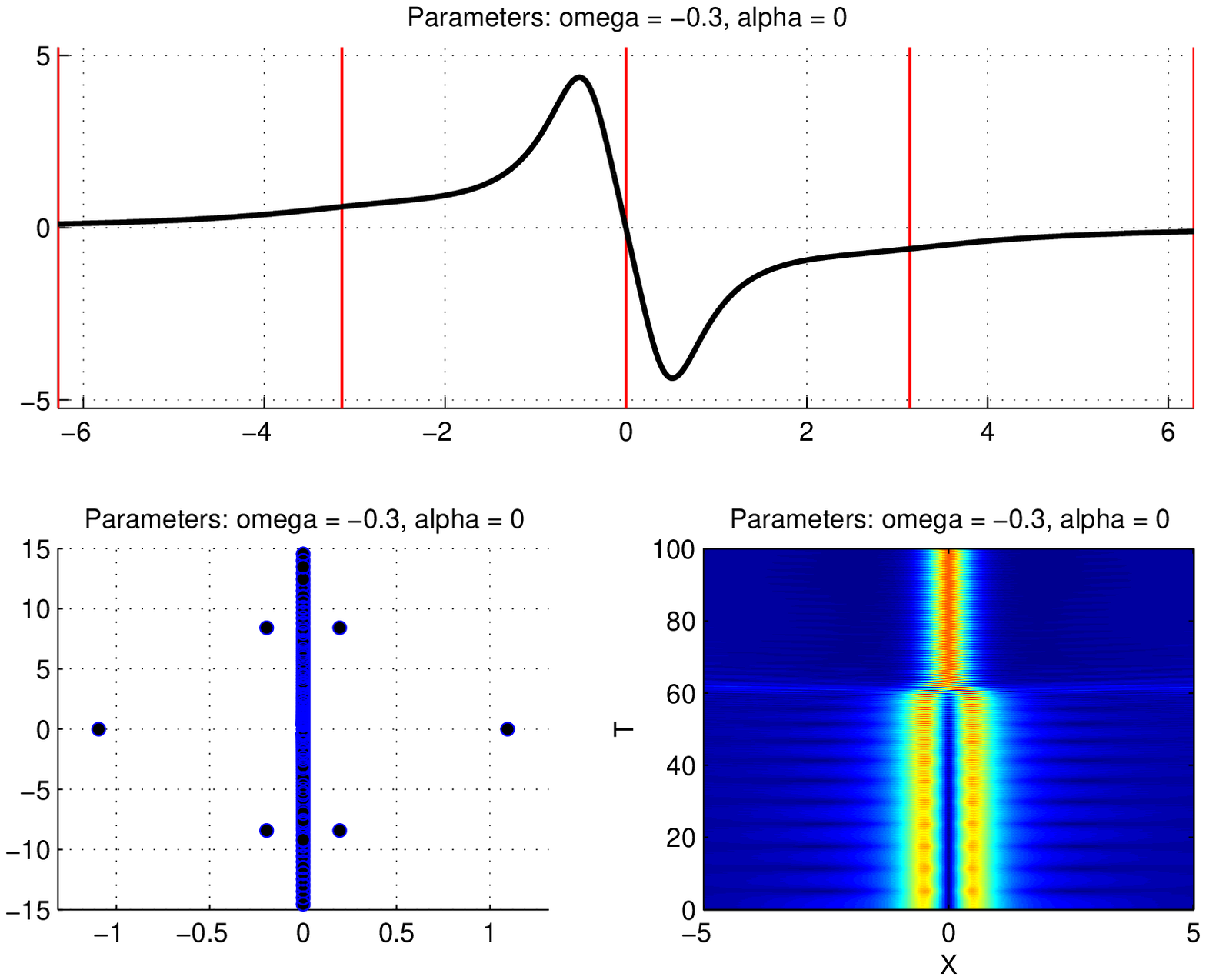}} %
\subfloat[][$\omega =
-0.7$]{\includegraphics[width=0.5\textwidth]{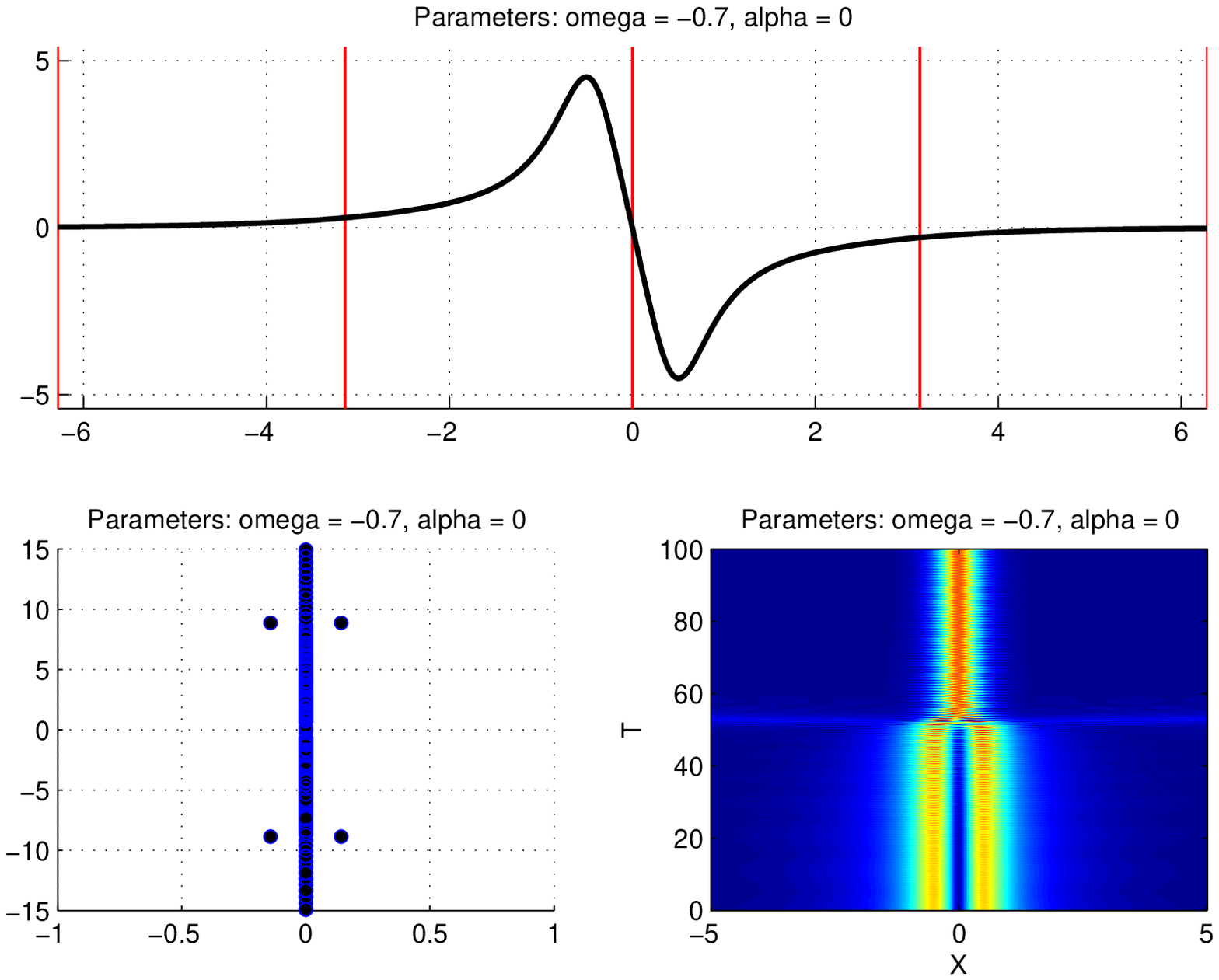}} \vfill
\subfloat[][$\omega =
-1.2$]{\includegraphics[width=0.5\textwidth]{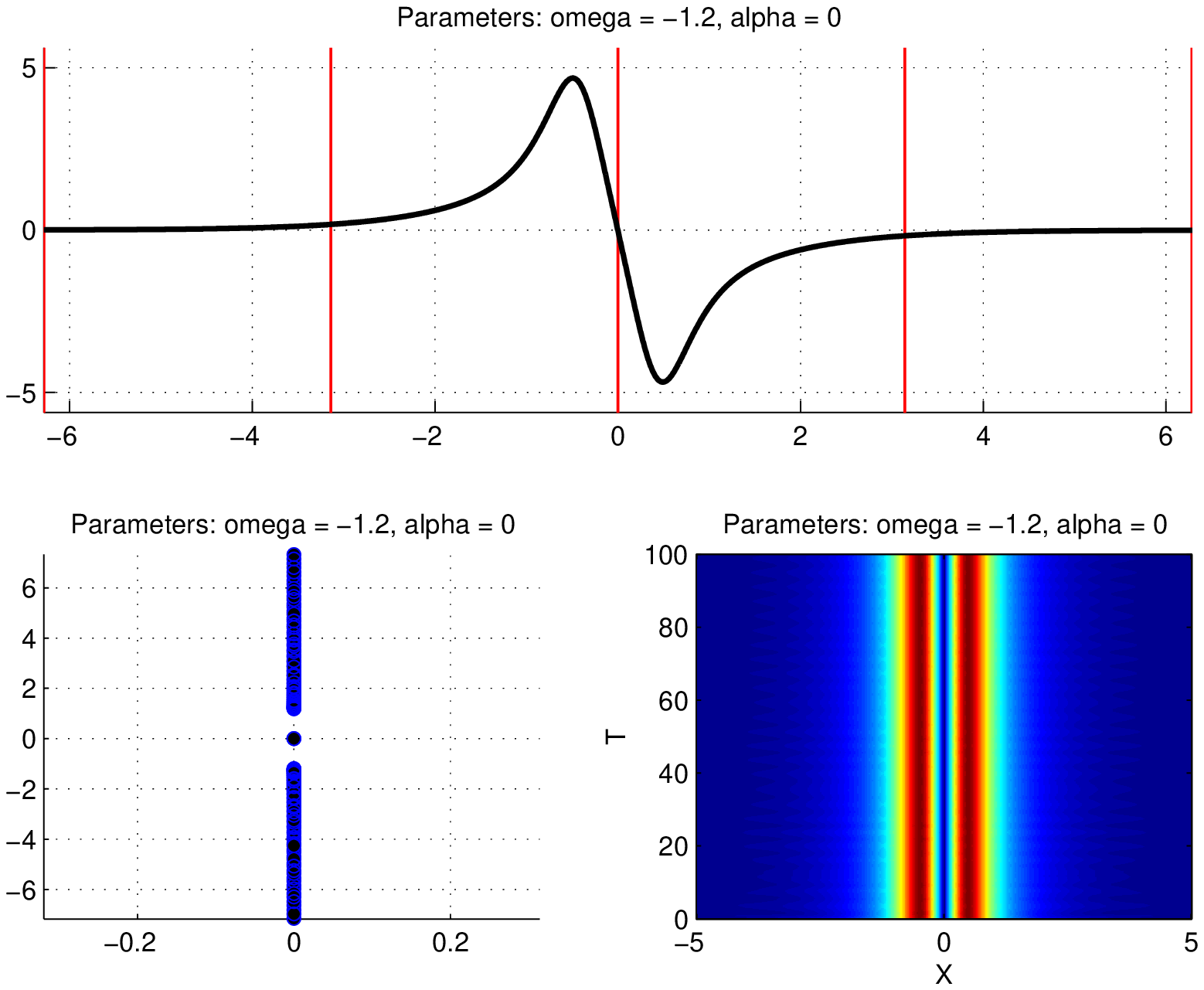}}  %
\subfloat[][$\omega =
-1.4$]{\includegraphics[width=0.5\textwidth]{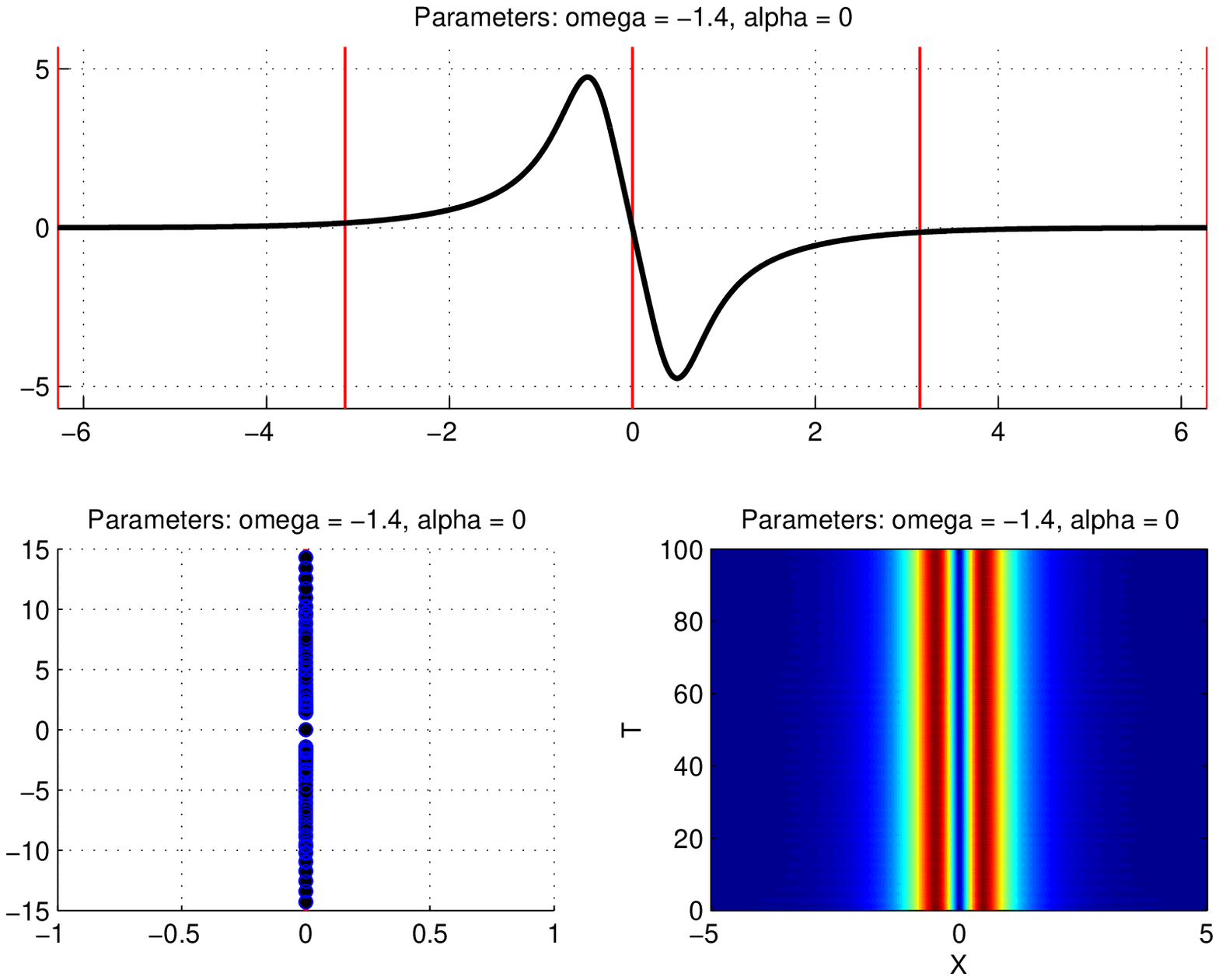}}
\caption{Typical examples of dipole solitons, their linear-stability
spectra, and unstable and stable temporal evolution, for $\protect\alpha =0$
in Eq. (\protect\ref{eq:current}). Additional examples of the evolution are
shown below in the lower panel of Fig. \protect\ref{pic:N(omega)}.}
\label{pic:sfs}
\end{figure*}

%\twocolumngrid

\begin{figure}[tbp]
\center{\includegraphics[scale=0.4]{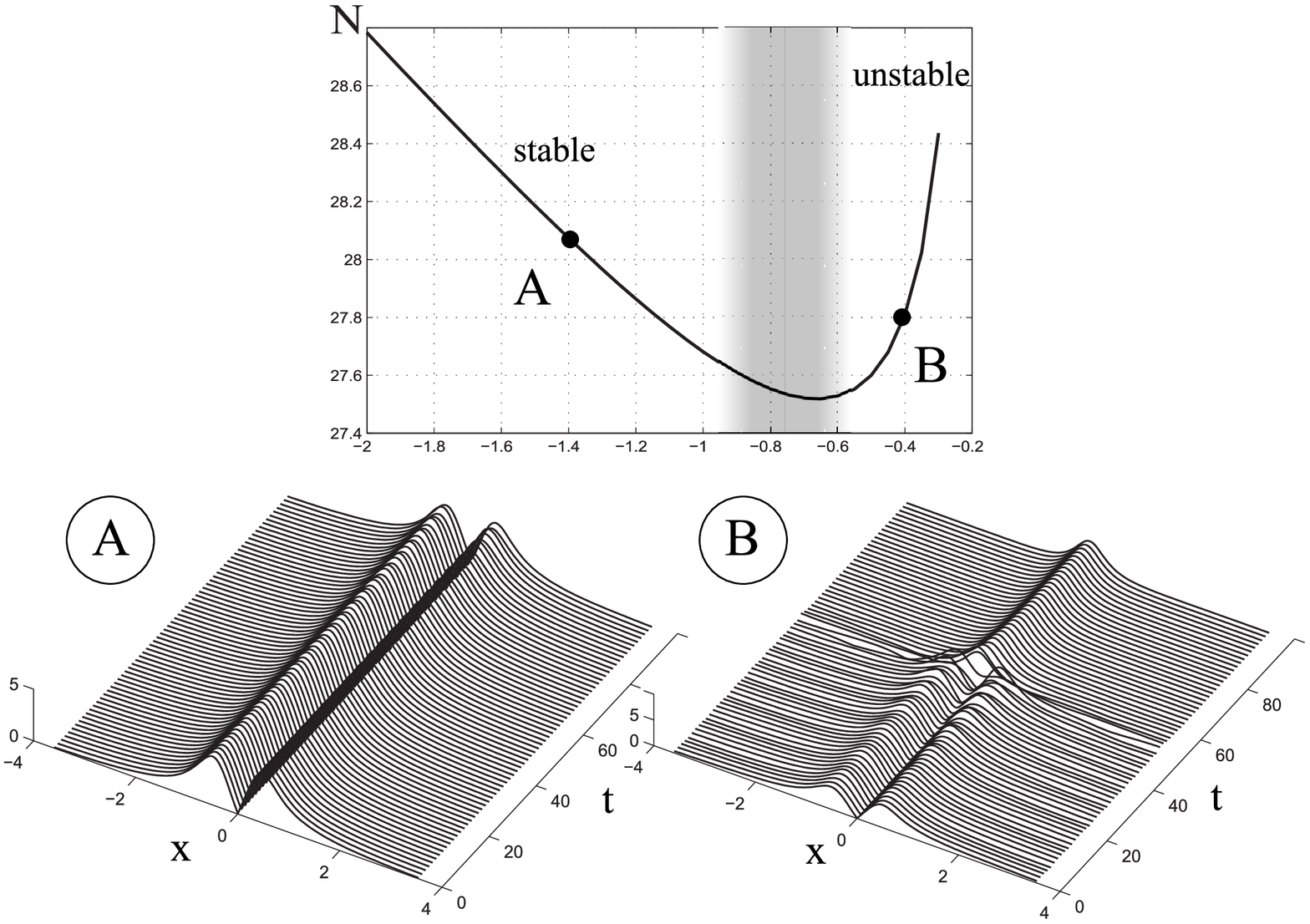}}
\caption{The upper panel: the $N(\protect\omega )$ curve for the
dipole solitons at $\protect\alpha =0$ in Eq. (\protect\ref%
{eq:current}). The lower panel displays typical examples of the stable and
unstable evolution of the dipole solitons for $\protect\omega =-1.4$
and $\protect\omega =-0.4$, respectively. In the latter case, the unstable
dipole soliton transforms into a fundamental soliton corresponding
to $\protect\omega \approx -13.96$ and amplitude $\approx 5.43$. In ``fuzzy area''
the simulation is very sensitive to the type of initial perturbation and the parameters of
the numerical method.}
\label{pic:N(omega)}
\end{figure}

Typical results of the simulations are presented in Fig. \ref{pic:sfs}, for $%
\alpha =0$ in Eq. (\ref{eq:current}). One can conclude that the VA\
prediction (iii), based on the VK criterion, is generally valid. The results are
summarized in the $\left( \omega ,N\right) $ plane, as shown in Fig. \ref%
{pic:N(omega)}. The DS is stable for the values of omega corresponding to
the slope of the $N(\omega)$ curve situated left to the minimum point $%
\omega _{\min }\approx -0.66$ and transforms into FS at the slope right to
this point. The border between the stability and instability regions in the
top panel of Fig. \ref{pic:N(omega)} is fuzzy. Within this ``fuzzy area''
the evolution of initial DS profile strongly depends on the type of imposed
perturbation and parameters of the numerical method.

\section{Conclusion}

\label{Discuss}The mathematical issue considered in this work is the
classification of families of solitons and their bound states in the model
of the nonlinear lattice, which is represented by the periodically varying
nonlinearity coefficient. A condition necessary for the existence of the
infinite variety of the bound states is that the local coefficient must
assume both positive and negative values. Then, the analysis is performed
for the physically relevant problem, which may find direct applications to
Bose-Einstein condensates and planar waveguides in nonlinear optics: finding
two branches of the DSs (\textit{dipole solitons}), whose
antisymmetric profile is confined, essentially, to a single cell of the
nonlinear lattice. The shape of these solitons is very similar to that of
the \textit{subfundamental solitons}, which are known in models with usual linear
lattice potentials, where they are chiefly unstable. An essential
finding reported here is that one of two branches of the single-cell DS, family which satisfies
the VK (Vakhitov-Kolokolov) criterion, is \emph{completely stable}. Also it was found that DSs belonging to the unstable branch evolve into stable FSs.
These results were obtained by means of numerical methods and also, in a
qualitatively correct form, with the help of the VA (variational
approximation). Besides that, it was found that particular species of FS
bound states are stable too.

\end{document}